\documentclass[aps.pre,showpacs]{revtex4}
\usepackage{graphicx}
\usepackage{color}

\def\be{\begin{equation}}
\def\ee{\end{equation}}     
\def\bfi{\begin{figure}}
\def\efi{\end{figure}}
\def\bea{\begin{eqnarray}}
\def\eea{\end{eqnarray}}

\begin{document}

\title{Equilibrium structure and off-equilibrium kinetics of a magnet with tunable frustration}
\author{Federico Corberi}
\affiliation {Dipartimento di Fisica ``E.~R. Caianiello'', and INFN, Gruppo Collegato di Salerno, and CNISM, Unit\`a di Salerno, Universit\`a  di Salerno, 
via Giovanni Paolo II 132, 84084 Fisciano (SA), Italy.}

\author{Manoj Kumar}

\affiliation{School of Physical Sciences, Jawaharlal Nehru University, New Delhi 110067, India.}

\author{Eugenio Lippiello}
\affiliation{Dipartimento di Matematica e Fisica, Seconda Universit\`a di Napoli,
Viale Lincoln, Caserta, Italy.}

\author{Sanjay Puri}

\affiliation{School of Physical Sciences, Jawaharlal Nehru University, New Delhi 110067, India.}

\begin{abstract}
 We study numerically a two-dimensional random-bond Ising model where
 frustration can be tuned by varying the fraction
$a$ of antiferromagnetic coupling constants. At low temperatures the model 
exhibits a phase with ferromagnetic order for sufficiently small values of $a$, $a<a_f$. 
In an intermediate range 
$a_f<a<a_a$ the system is paramagnetic, with spin glass order expected right at zero temperature.
For  even larger values $a>a_a$ an antiferromagnetic phase exists. 
After a deep quench from high temperatures, slow evolution is observed for any value of 
$a$. We show that different amounts of frustration, tuned by $a$, affect the dynamical 
properties in a highly non trivial way. In particular, the kinetics is logarithmically slow in phases 
with ferromagnetic or  antiferromagnetic order, whereas evolution is faster, i.e. algebraic,
when spin glass order is prevailing. An interpretation is given in terms of the different nature
of phase space. 

\end{abstract} 


\maketitle

\section{Introduction} \label{intro}

When a system is brought across a phase-transition towards a state where the initial symmetry
is spontaneously broken, a slow non-equilibrium evolution sets in. A paradigm is represented 
by binary systems~\cite{bray}, such as ferromagnets, cooled from above to 
below the critical temperature $T_c$. In the clean case, namely
in the absence of any kind of quenched disorder or inhomogeneities, the kinetics is characterised by a coarsening 
process that is nowadays quite well understood. Domains of the symmetry-related equilibrium
phases at the final temperature $T_f$ form and their typical size $L(t)$ grows in time.
This process is endowed with a scaling symmetry such that configurations visited
at different times are statistically equivalent provided lengths are measured in 
units of $L(t)$. The growth law of the domains' size is generally algebraic $L(t)\sim t^{1/z}$. 
The dynamical exponent $z$ is temperature independent and varies only among different  dynamical 
universality classes which -- in turn -- are determined by the symmetry of the order parameter, i.e. if scalar 
or vectorial, and by the character of the dynamics, e.g. in presence of conservation laws, hydrodynamics 
etc...

Phase-ordering can occur in disordered ferromagnets as well. In these systems an amount
of quenched randomness is present, but its effects are sufficiently weak not to spoil the basic structure
of the equilibrium state. A disordered phase at high temperature and a low temperature
one are still present and the symmetry breaking mechanism is akin to that of clean magnets -- 
e.g. up-down (or $Z_2$) symmetry is spontaneously broken for a scalar order parameter. 
Examples are magnetic models in the presence of
random external fields, coupling constants with a stochastic component, or quenched vacancies~\cite{crp}.

If disorder is sufficiently weak not to change the equilibrium structure, its presence is by far much
more important as dynamical properties are concerned.  Indeed one observes that even the smallest
amount of randomness usually slows down dramatically the asymptotic growth law of the domains.
This is because
quenched disorder pins the dynamics introducing energetic barriers which can only be 
exceeded by rare thermal fluctuations. Therefore, at variance with the clean systems, the dynamical features are 
strongly temperature dependent. 

While equilibrium properties are sometimes quite understood, thanks also to
some general results such as the Harris criterion, understanding of the off-equilibrium evolution
is by far incomplete. Concerning the growth law of the domains' size, either logarithmic or 
temperature dependent power-laws have been reported both in experiments~\cite{exprf,exprb,exprb2,exprb3}
and in model systems~\cite{maz,grsr,rao2,LipZan10,Fisher,Decandia02,rflog3,noirf,rflog2,sanhi,otherfavorSU2,
dil1,dil2,HH,villain,otherfavorSU1,henkelrb,noi_rb,noi_diluted,noi_3d,rough_coars,insalata}, and there is yet no
clear indication of a simple classification -- e.g. on the basis of some dynamical universality classes --
of the behavior of different disordered magnets. Moreover, despite the fact that disorder is responsible of 
pinning and slowing down of the kinetics, the simple idea that the more disorder is present in a system
the slower the evolution will be, has been recently shown~\cite{noi_rb,noi_diluted,noi_3d} to be incorrect in quite a number of cases.
Considering for instance the Ising model with random dilution  (namely a fraction $d$  of sites or bonds  
on the lattice are missing) it was shown that, although for sufficiently small values of $d$ the kinetics is 
slowed down upon increasing $d$, as naively expected, 
after a certain threshold increasing $d$ produces a faster growth. As it is explained
in~\cite{noi_diluted, noi_rb}, this happens because adding more disorder -- in this case parametrised by 
$d$ -- not only one introduces more pinning sources but, more importantly, the topological properties
of the system are also changed. Indeed, when $d$ gets close to the value $d_c$, where the set of non-diluted
sites (or bonds) is at the percolation threshold, the fractal properties of the network play an important role
in speeding up the evolution, because the pinning barriers are softened.

This very effect, the non monotonous behavior of the speed of growth vs
the amount of disorder, is observed not only in diluted systems but also for Ising spins
with random ferromagnetic couplings~\cite{noi_rb}, the kind of model that will be generalised 
with the addition of frustration in this Article.
  
All the systems described insofar are non-frustrated. 
When it is impossible to simultaneously satisfy all the interactions between the microscopic constituents,
as in the paradigmatic example of the antiferromagnetic Ising model on the triangular lattice,
frustration arises.
The slow evolution of disordered frustrated systems 
is by far a much more complicated problem. This is because even the basic structure of the low temperature
equilibrium states in finite-dimensional systems is still debated. The absence of a clear-cut indication on the
static properties hinders
the interpretation of what is dynamically observed. Indeed, in a droplet theory scenario one would expect
a kind of coarsening reminiscent of what previously discussed for ferromagnetic systems,
while if a picture inspired to the mean-field solution applies, something very different should happen
\cite{yosh}.

The aim of this paper is to gradually near the study of the kinetics of disordered systems with frustration 
from the side of non-frustrated ones, where a better understanding has been to some extent achieved.
In order to do that we consider an Ising model with a fraction $a$ of antiferromagnetic coupling constants, 
and the remaining ones are ferromagnetic. We study the model numerically in two dimensions as the
value of $a$ is varied in $[0,1]$. Clearly, by changing the parameter $a$ one can gradually tune the amount of frustration present in the
system. Not enough, in order to soften as possible the crossover from a situation without frustration to 
one where it is relevant, we consider the case where the ferromagnetic interactions are much stronger than the antiferromagnetic ones. This allows us
to stay as close as possible -- so to say -- to the simpler and better 
understood ferromagnetic situation.

In the low-temperature equilibrium phase-diagram of the model, as $a$ is progressively increased,
one moves from a ferromagnetic phase, where frustration plays a minor role, to a strongly frustrated 
paramagnetic phase which, right at $T=0$, is expected~\cite{jorg96} to exhibit spin glass order~\cite{Mezard}.
For even larger values of $a$ an antiferromagnetic region is entered. 
Therefore, considering the evolution of the present model after a deep quench 
to a small finite temperature $T_f>0$,
 one has the opportunity to study how different amounts of frustration influence the off-equilibrium
 kinetics. In doing that we find that in the magnetic phases -- either ferromagnetic or 
 antiferromagnetic -- a usual coarsening is observed characterised by a logarithmic increase
 of the domains' size $L(t)$, in agreement with previous studies on related Ising models with
 random bonds~\cite{noi_rbpowerlog}.  
Upon increasing frustration the speed of 
phase-ordering changes in a non-monotonous way. This behavior, which is analogous to the 
one discussed above 
 for non-frustrated systems, can be interpreted along similar arguments based on topology. 
Indeed the geometry of the growing domains becomes fractal as $a$ is increased and the
transition to the paramagnetic region is approached, similarly to what happens in the 
non-frustrated diluted systems previously considered when the percolation threshold is approached.
Also in this case the fractal topology speeds up the evolution.

 This shows that an off-equilibrium evolution getting faster and more efficient  with the addition 
 of disorder is of a quite general nature, and occurs both in system with and without frustration. 
At variance with the logarithmically slow evolution observed in the ferro or 
antiferromagnetic phases, a faster kinetics characterised by algebraic behaviors  
 is found along the whole paramagnetic region, 
where frustration plays a prominent role. In this region, although neither ferromagnetic, antiferromagnetic or
spin glass order~\cite{Mezard} are present at finite temperatures, 
a quench to $T_f>0$ exhibits slow evolution as due to the
proximity of the spin glass ground state at $T_f=0$~\cite{jorg96}. 
Due to that, the faster evolution observed in this phase (as compared to the logarithmic one in the ferro and
antiferromagnetic regions) can be perhaps ascribed
to the spin glass structure with many quasi isoenergetic levels and softer
barriers as opposed to those present in a ferromagnetic phase with two profound
free energy minima.

This paper is organised as follows: In Sec. \ref{model} we introduce the model, set the notation,
and discuss the structure of the bond network in Sec. \ref{geom_bonds}.
The next section, Sec. \ref{ph_diag}, is devoted to the study of the low-temperature phase-diagram of the system.
The results concerning the kinetics of the model after deep quenches to various final temperatures 
are presented and discussed in Sec. \ref{sec_kin}. Finally, we summarise and draw our conclusions in the last Section.

\section{The model} \label{model}

We consider the Ising model with Hamiltonian
\be
{\cal H}\left (\{s_i\}\right )=-\sum _{\langle ij\rangle}J_{ij}s_is_j
\ee
where the $s_i=\pm 1$ are spin variables, the sum runs over nearest neighbours
couples $\langle ij\rangle$ of a lattice, and the coupling constants
are $J_{ij}=J_0+\xi_{ij}$, where $J_0>0$ and the $\xi_{ij}$ are 
uncorrelated random variables extracted from a bimodal distribution
\be
P(\xi )=a\,\delta _{\xi,-K}+(1-a)\,\delta _{\xi,K},
\ee
where $\delta$ is the Kronecker function and $0\le a\le 1$ and $K$ are parameters.
In a previous paper~\cite{noi_rb} we studied the kinetics of this model without frustration,
with $K <J_0$,  namely with only ferromagnetic interactions.
Here instead we set 
\be
K>J_0,
\label{condanti}
\ee
meaning that the fraction $a$ of bonds with 
$\xi _{ij}<0$ are antiferromagnetic and the remaining ones are
ferromagnetic. We will also denote with $J_-=J_0-K$ and $J_+=J_0+K$ the strength of 
such bonds, respectively. 
We will consider a square lattice in $d=2$ with periodic boundary conditions.

Next to the ferromagnetic local order parameter
$s_i$, the spin, it is useful to introduce the antiferromagnetic one
\be
\sigma _i=(-1)^i s_i,
\label{stagspin}
\ee
or {\it staggered} spin. In Eq. (\ref{stagspin}) it is stipulated that the index $i$ runs over the lattice sites in
such a way that two nearest neighbors always have an opposite value of $(-1)^i$.
 
A classification of the low-temperature equilibrium states will be made in Sec. \ref{ph_diag} in terms of their
spontaneous magnetisation 
\be
m=\frac{1}{N} \sum _is_i,
\ee
and of the staggered magnetisation
\be
M=\frac{1}{N} \sum _i \sigma_i.
\ee

\subsection{The geometry of the bond network}\label{geom_bonds}

\begin{figure}[h]
  \centering
  \rotatebox{0}{\resizebox{.5\textwidth}{!}{\includegraphics{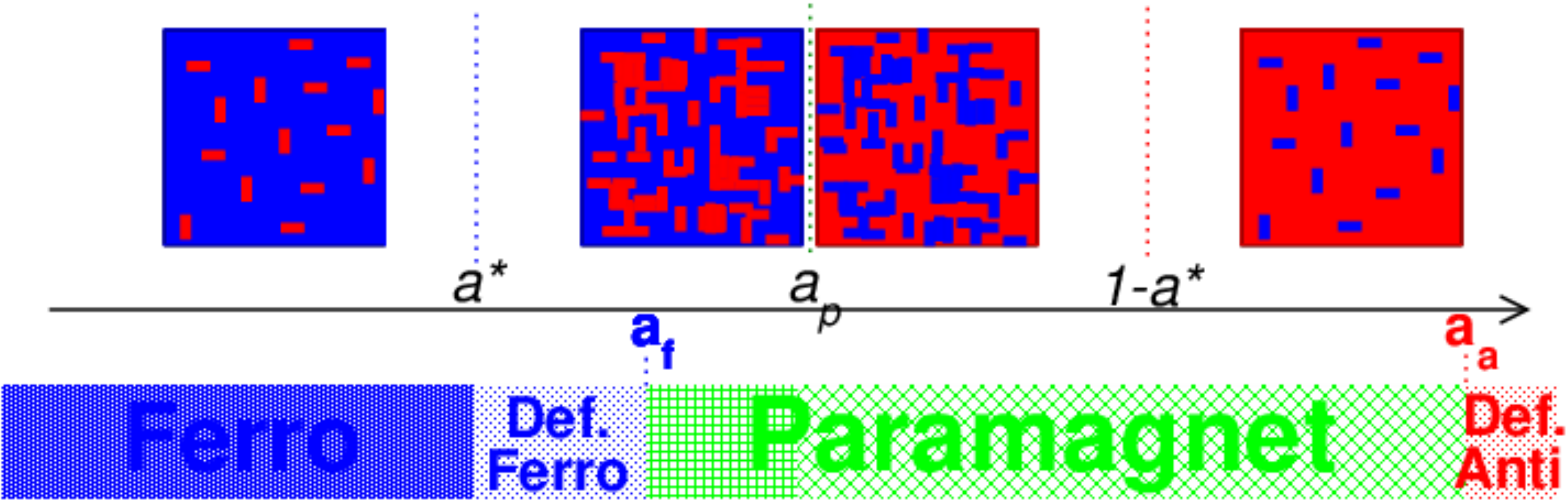}}}

    \vspace{2cm}
  
  \rotatebox{0}{\resizebox{.5\textwidth}{!}{\includegraphics{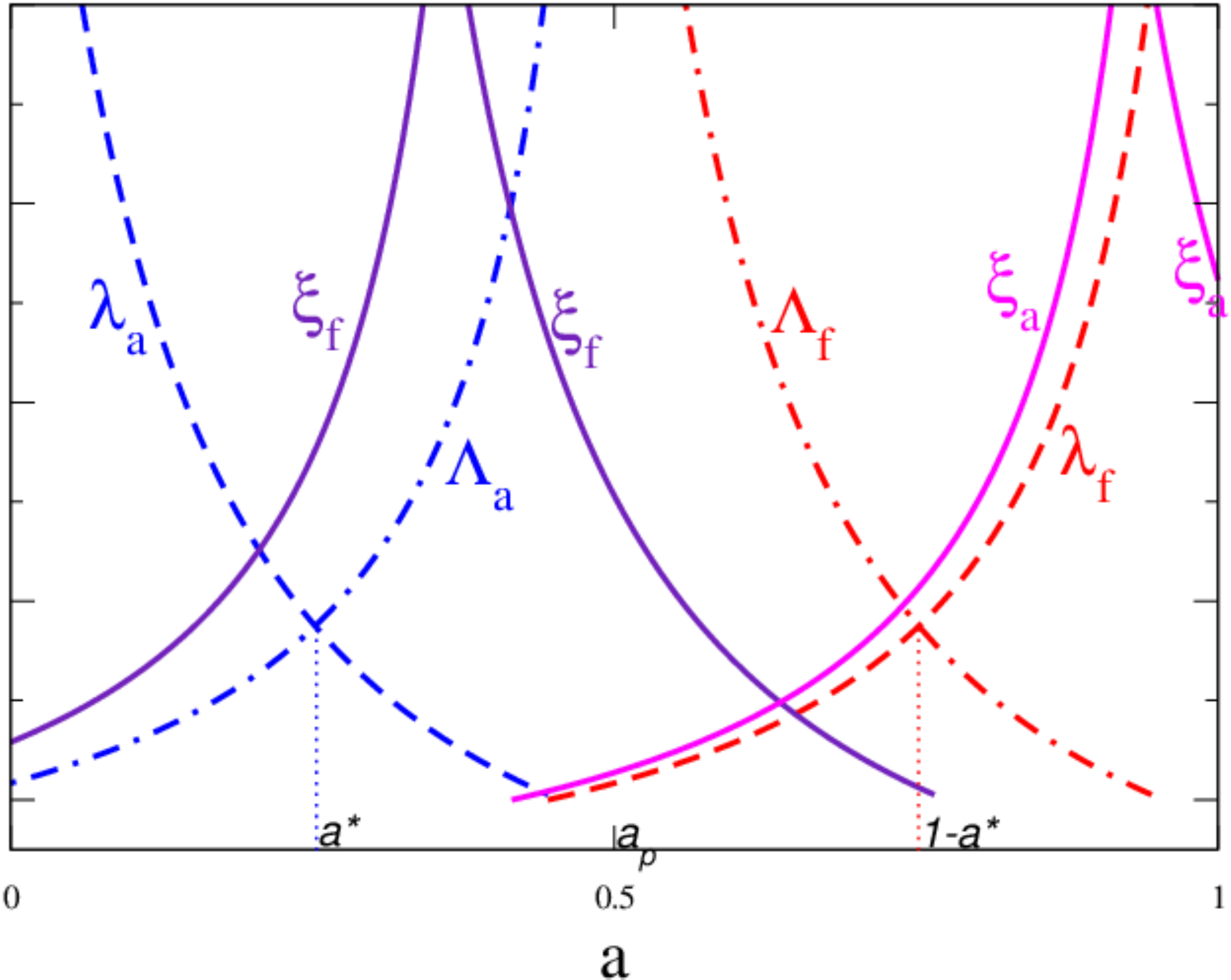}}}
  \caption{In the upper stripe four typical bond configurations are pictorially shown,
    corresponding to $0<a<a^*$, $a^*<a\lesssim a_p$, $a_p\lesssim a<1-a^*$ and
    $1-a^*<a<1$, from left to right respectively.
    Ferromagnetic bonds are drawn in blue, antiferromagnetic ones in red.
    The bar below the configuration stripe describes the physical phases of the
    systems as $a$ is varied, e.g. if ferromagnetic, paramagnetic etc.
    The graph in the lower part of the figure is a schematic representation
    of the behavior of the typical lengths characterising the bond configuration
    and the physical properties (see text).}
\label{fig_substrate}
\end{figure}

It is useful to discuss the geometrical properties of the network of
bonds of the model, which is pictorially illustrated in the upper stripe of
Fig.~\ref{fig_substrate}. With $a=0$ the system is the usual clean
Ising ferromagnet, since all the coupling constants are positive.
Moving to a finite $a$ amounts to add some antiferromagnetic bonds.
If $a$ is small, these will be separated apart by a typical
distance 
\be
\lambda _{a}\sim a^{-1/d}.
\label{lambdaa}
\ee
This situation is schematically represented in the leftmost box on the 
upper stripe of the figure. Here the blue colour corresponds to regions where
the bonds are ferromagnetic while antiferromagnetic ones are drawn in red.
The behavior of $\lambda _{a}$ is shown by a dashed blue line 
in the lower graph of Fig.~\ref{fig_substrate}.
It decreases as $a$ raises, meaning that at some point 
it becomes of the order of the lattice spacing and groups of
antiferromagnetic bonds start to coalesce. Indeed, we know that for
$a=a_p=1/2$ such bonds form a percolating cluster, since 
the bond percolation threshold is precisely $a_p$. 
Due to this,
right at $a_p$ the size $\Lambda _{a}$ of the regions of clustered
antiferromagnetic bonds is $\Lambda _{a}=\infty$ and, for
$a$ smaller but not to far from $a=a_p$ one has the well known percolative behavior
\be
\Lambda _{a}=(a_p-a)^{-\nu},
\ee
with $\nu =4/3$, which is shown by a dotted-dashed blue line in 
the lower part of Fig.~\ref{fig_substrate}. A pictorial representation of the bond
configuration in the region $a\lesssim a_p$ is shown in the 
second (from the left) box in the upper part of the figure.
The transition between the region with isolated and clustered 
antiferromagnetic bonds occur at a value $a=a^*$ which can be
roughly identified as the point where $\lambda _{a}\simeq \Lambda _{a}$.
The actual value of $a^*$ is located~\cite{noi_rb} between $a=0.2$ and $a=0.3$.

Clearly, as we move in the region $a>a_p$ the situation mirrors
that for $a<a_p$ upon exchanging the roles of ferromagnetic and 
antiferromagnetic bonds and substituting the lengths $\lambda _{a},\Lambda _{a}$
with the corresponding ones $\lambda _{f},\Lambda _{f}$ relative to the 
ferromagnetic bonds, for which one has
\be
\lambda _{f} =(1-a)^{-1/d},
\label{lambdaf}
\ee
and
\be
\Lambda_{f}=(a-a_p)^{-\nu}.
\ee

\vspace{5cm}

\section{The structure of the equilibrium states} \label{ph_diag}

When $J_0>0$ the symmetry of the 
bond geometry for $a<a_p$ and $a>a_p$ is spoiled because the actual 
absolute value of positive and negative bonds is different.  
Furthermore, the simultaneous presence of positive and negative
couplings may introduce frustration and make the system highly non trivial
as in the notable case of a spin glass, which can be obtained in the present 
model by letting $J_0=0$. 
In this paper we focus on the somewhat simpler case
with
\be
K <\frac{z}{z-2}J_0,
\label{condition}
\ee
where $z$ is the coordination number of the lattice.
Eq. (\ref{condition}) is a {\it ferromagnetic-always-wins} condition.
Indeed it can be simply checked that, when Eq. (\ref{condition}) holds,
a spin to which at least a ferromagnetic bond is attached will
always lower its energy by pointing along the direction of the majority
(if a majority exists) of spins to which it is connected by 
ferromagnetic bonds. For instance, a spin attached to a single ferromagnetic
bond will decrease its energy by aligning with the spin
on the other side of that bond, irrespective of the configuration
of the other $z-1$ neighboring spins.  
Still being frustrated, a system conforming to the condition (\ref{condition}) 
has a simpler structure -- although not trivial at all -- of the low-temperature equilibrium states,
as we will see shortly. Notice also that the 
spin glass do not obey Eq. (\ref{condition}) (since $J_0=0$).
All the numerical data that will be presented in the following refer to the case 
$J_0=1$, $K=5/4$, which obviously obeys Eq.  (\ref{condition}).

Let us now discuss the structure of the low-temperature equilibrium states of the model
as $a$ is changed. 

\subsection{Ground states}

In the following we will focus on the ground states, namely the equilibrium configurations
at $T=0$. We found these states by a generalisation~\cite{4collegueUk} to systems with
periodic boundary conditions of the algorithm introduced in~\cite{2collegueUk}. 
With this technique the ground state can be found in polynomial time.
We will postpone the discussion of the effects of a finite temperature to Sec. \ref{eq_finiteT}.

\begin{figure}[h]
\vspace{2cm}
\centering
\rotatebox{0}{\resizebox{.65\textwidth}{!}{\includegraphics{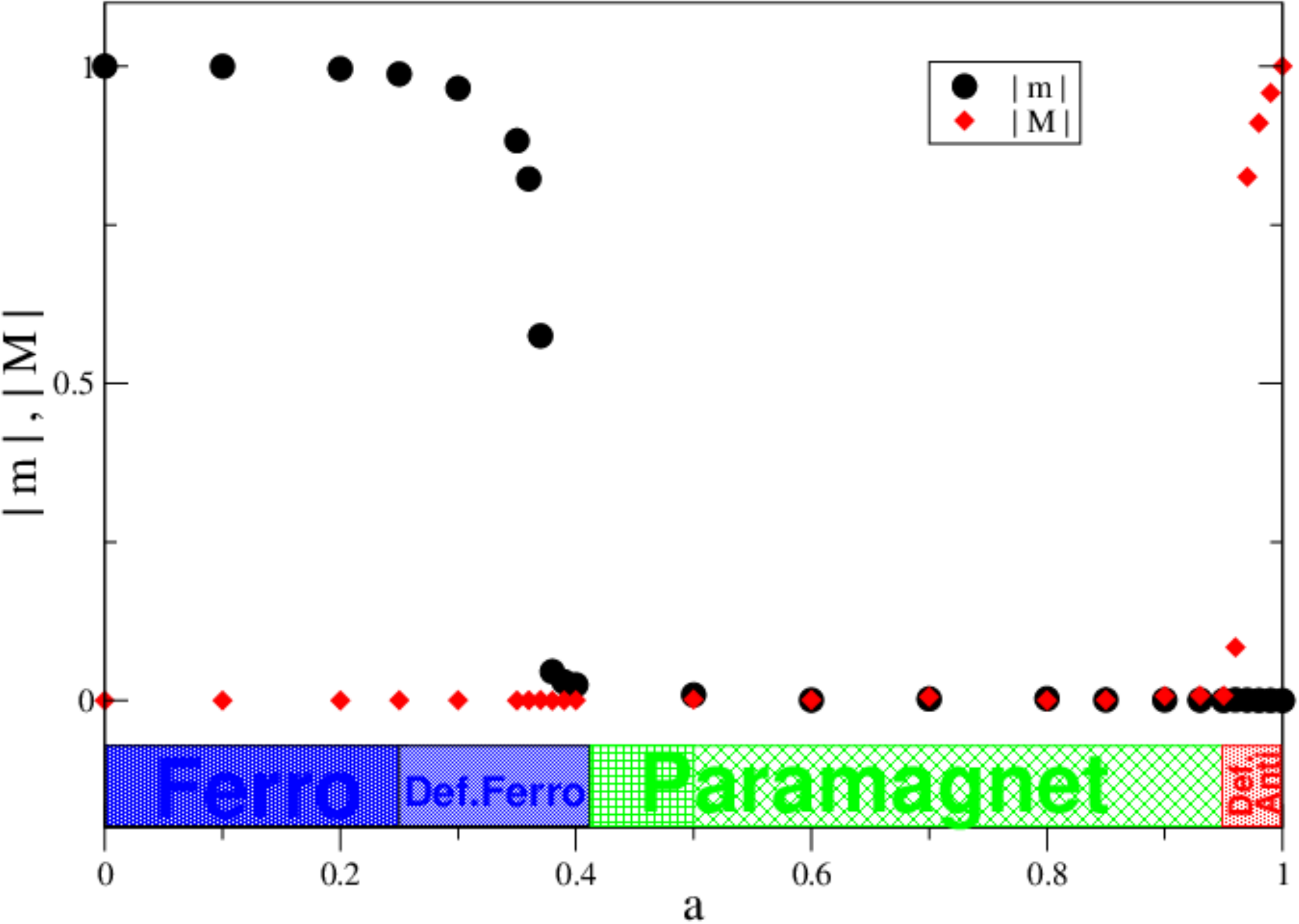}}}
\caption{$|m|$ and $|M|$ at $T=0$ for different values of $a$.}
\label{fig_magn}
\end{figure}

The global order parameters $m$ and $M$ defined above, computed in the ground states 
of the model for different values of $a$, are shown in Fig.~\ref{fig_magn}. 
We will classify the ground states according to $m$ and $M$ in the 
different regions of the parameter $a$ in the following sections. This classification is reported in the upper bar of Fig.~\ref{fig_substrate} and, similarly, in the lower bar of Fig.~\ref{fig_magn}.

\begin{figure}[h]
\vspace{2cm}
\centering
\rotatebox{0}{\resizebox{.95\textwidth}{!}{\includegraphics{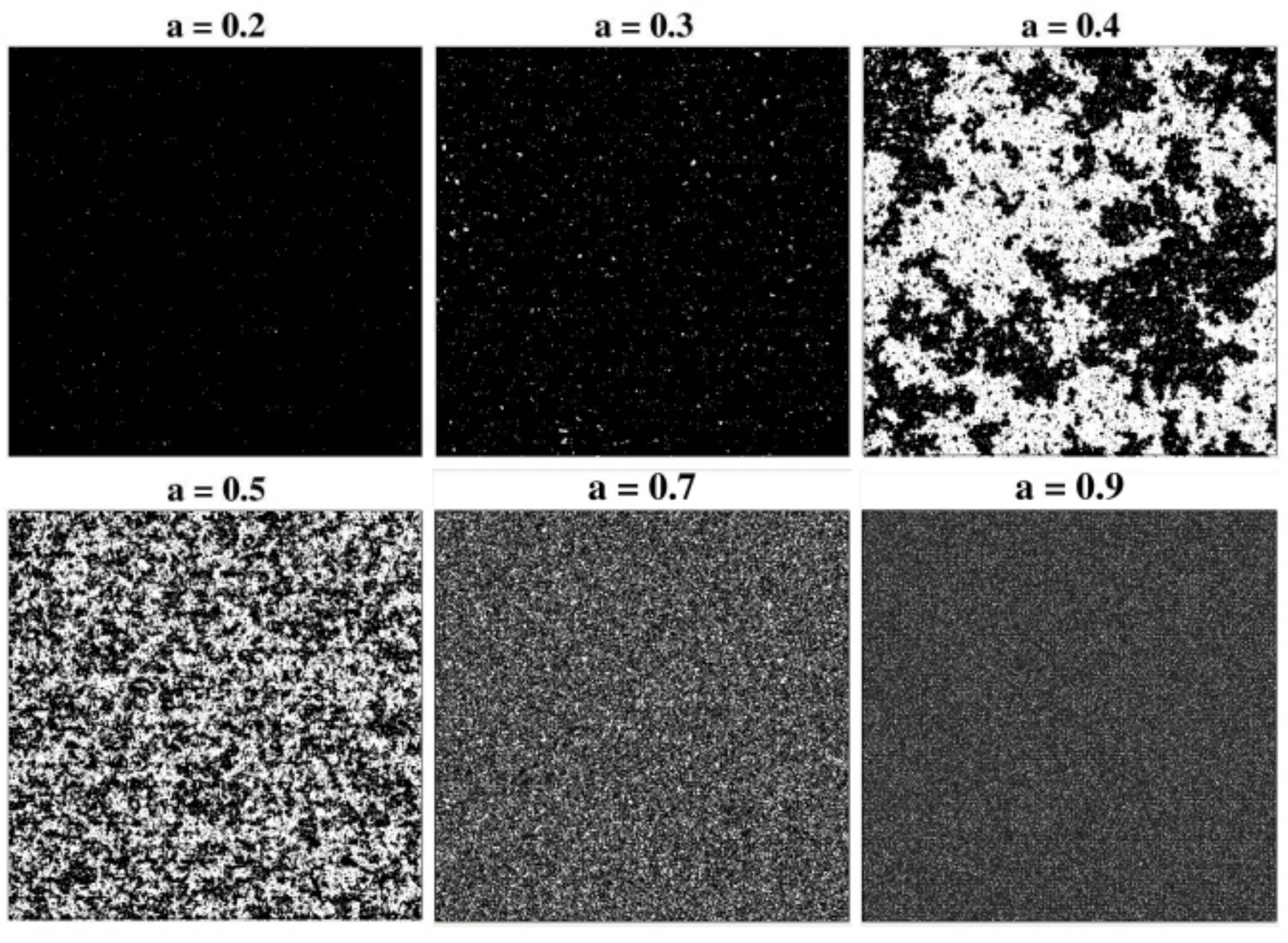}}}
\caption{Configurations of the ground state for a system of size ${\cal L}=512$ 
for different values of $a$. Spins up are plotted in black, spins down in white.}
\label{fig_gs}
\end{figure}

\begin{figure}[h]
\vspace{2cm}
\centering
\rotatebox{0}{\resizebox{.95\textwidth}{!}{\includegraphics{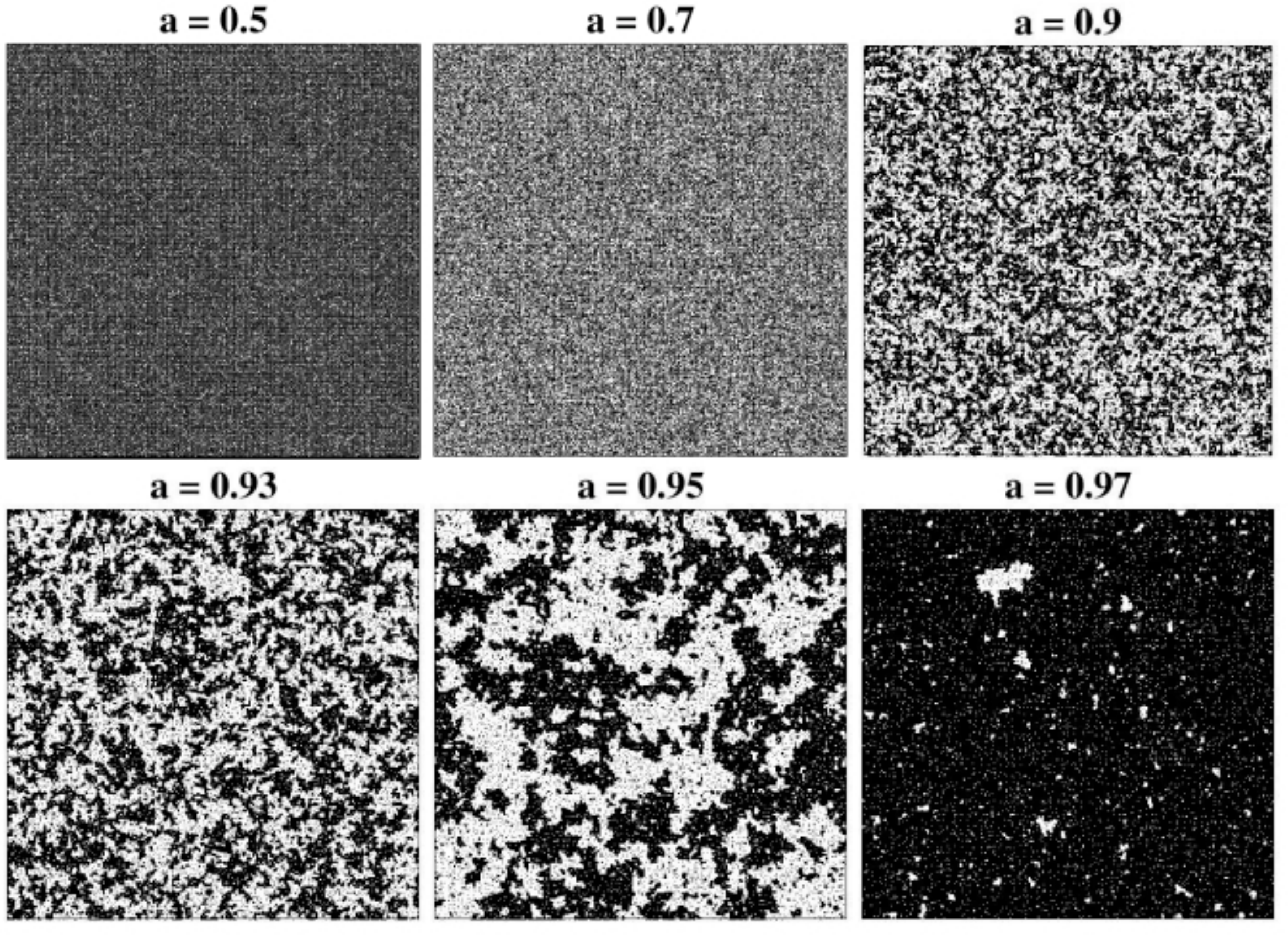}}}
\caption{Configurations of the ground state for a system of size ${\cal L}=512$ 
for different values of $a$. We plot the staggered spin $\sigma _i$ defined in Eq. (\ref{stagspin}).
$\sigma _i=1$ are plotted in black, $\sigma _i=-1$ in white.}
\label{fig_gs_anti}
\end{figure}

\subsubsection{Ferromagnetic ($0\le a<a_f$)} \label{reg1}

Let us start discussing the ferromagnetic phase, which can be split in
the two sectors with $0\le a<a^*$ and $a^*\le a <a_f$, that will be considered
separately below.
\vspace{.5cm}

{\it Sector } $0<a<a^*$
\vspace{.3cm}

As discussed above in Sec. \ref{geom_bonds}, 
in this region there are basically only isolated
antiferromagnetic links in a sea of ferromagnetic ones.
At $T=0$ spins in this sea necessarily align in a ferromagnetic state.
Condition (\ref{condition}) implies that
also the spins attached to the antiferromagnetic bonds must be aligned with 
those in the sea. Hence the ground state is akin to a usual ferromagnetic system.
Of course, as $a$ increases, there is a finite probability
to find some antiferromagnetic bonds nearby and this can cause some spin reversal 
with respect to a completely ordered configuration, but for $a<a^*$ these are 
quite few.
Since the presence of antiferromagnetic bonds is largely irrelevant in this parameter region
we expect $|m|\simeq 1$ and $M=0$. We see in Fig.~\ref{fig_magn}
that this is indeed the case.
 
A representation of a real ground state for $a=0.2<a^*$ 
(let us recall that $a^*$ is expected to be located between $a=0.2$ and $a=0.3$)
is shown in the upper left panel
of Fig.~\ref{fig_gs}, which confirms the description above 
\vspace{.5cm}

{\it Sector } $a^*\le a<a_f$
\vspace{.3cm}

Fig.~\ref{fig_magn} shows that ferromagnetic order extends up to
a certain $a=a_f$ (with $a_f \gtrsim 0.4$), which is well beyond $a^*$.
Ferromagnetic ordering occurring beyond $a^*$ does not come as a surprise,
since in the whole region $a<a_p$ the number of ferromagnetic bonds
is larger than that of the antiferromagnetic ones and also because of the 
{\it ferromagnetic-always-wins} condition (\ref{condition}).
Notice however that this does not guarantee that ferromagnetic order
is sustained up to $a=a_p$ or beyond, for the reasons that
will be explained in Sec. \ref{regpara}, but only up to a lower value $a=a_f<a_p$.

In the region $a^*<a<a_f$ there is still a prevalence of ferromagnetic order, namely
the fraction of -- say -- up spins prevails over the reversed ones,
but since antiferromagnetic bonds can coalesce, regions with down spins may be found
locally, as it can be seen in Fig.~\ref{fig_gs} for $a=0.3$ and $a=0.4$
(upper central and right panel). This is why we call this situation {\it defective ferromagnet}.
Clearly the islands where spins are reversed
increase upon raising $a$, as it can be checked in Fig.~\ref{fig_gs}.
Here one sees that the size $\xi _f$ of these regions 
grows dramatically as $a$ gets close to $a_f$, a fact that is pictorially sketched in Fig.~\ref{fig_substrate},
as in the presence of a continuous phase-transition.
The presence of extended regions opposite to the dominant order clearly depletes the 
magnetisation of the system, so when $a< a^*< a_f$ one has $0<|m|<1$ (decreasing upon 
raising $a$) and $M=0$, as it can be observed in Fig.~\ref{fig_magn}.
The magnetisation $m$ vanishes at the transition point $a=a_f$.

\subsubsection{Paramagnetic phase ($a_f\le a\le a_a$)}\label{regpara}

We discuss separately the two sub-regions with $ a_f<a<a_p$ and $a_p<a<a_a$ below.

\vspace{.5cm}

{\it Sector } $ a_f<a<a_p$
\vspace{.3cm}

Here the ferromagnetic bonds still prevail and form a sea which spans the
system. The difference
with the ferromagnetic region is that antiferromagnetic bonds, besides being
grouped together, can form sufficiently connected paths as
to destroy the ferromagnetic state. This is discussed in Appendix \ref{app1}.

A real configuration of the system in this region looks like the one for $a=0.5$ in Fig.~\ref{fig_gs}
(bottom raw, left). Notice also that, upon comparing this configuration with the one at $a=0.7$ one sees
that the size $\xi _f$ of the locally magnetised regions increases as $a$ decreases towards $a_f$,
suggesting that $\xi _f$ diverges also on this side of $a_f$, as it is sketched in Fig.~\ref{fig_substrate}.

Given the structure of the ground state discussed above, one has $m=0$. Clearly it is also
$M=0$, since negative bonds are a minority and there cannot be antiferromagnetic 
ordering. This is confirmed in Fig.~\ref{fig_magn}.
For this reason we generically denote the region with $a_f\le a \le a_a$
as {\it paramagnetic}.
Let us anticipate, however, that right at $T=0$ some spin glass order is expected,
as we will further discuss in Sec. \ref{paradyn}.

\vspace{.5cm}

{\it Sector } $a_p<a<a_a$
\vspace{.3cm}

In this region there is a sea of antiferromagnetic bonds. 
If $a$ is larger but sufficiently close to $a_p$ there are also ferromagnetic islands
inside which spins are aligned. However these islands are disconnected 
and hence they order incoherently. Therefore we expect $m=0$ throughout
the region $a>a_p$. This is observed in Fig.~\ref{fig_magn}.
The presence of a spanning sea of antiferromagnetic bonds is not
sufficient to guarantee that a global antiferromagnetic order will establish,
not even if $a$ is so large that ferromagnetic bonds are isolated,
which happens for $a>1-a^*$ (we recall that $a^*$ lies between $0.2$ and $0.3$).
Indeed we see in Fig.~\ref{fig_magn} that the property $M=0$ extends up to $a=a_a$, where
$a_a$ is located around $a\gtrsim 0.95$.
The development of antiferromagnetic order cannot be easily observed by representing
the value of the spins, as done in Fig.~\ref{fig_gs}, since on a large scale
one gets a uniform grey plot.
On the other hand, it can be clearly seen by plotting the staggered spin
$\sigma _i$ instead of $s_i$, as it is done in Fig.~\ref{fig_gs_anti}.
In this figure 
local antiferromagnetic order results in black or white regions, and $|M|>0$
corresponds to a majority of one of the two colours.

The very reason why antiferromagnetic order cannot establish up to such
large fraction of
antiferromagnetic interactions as $a=a_a$
is obviously related to the {\it ferromagnetic-always-wins} condition
(\ref{condition}), as it is discussed in Appendix \ref{app2}.
Beyond $a_a$ 
the antiferromagnetic order sets in. This region will be discussed below (Sec. \ref{reg4}).

It is nowadays quite well established~\cite{jorg96}
that for the two-dimensional spin glass model
(corresponding to the choice $J_0=0$ in our notation) a zero temperature
spin glass phase exists, which however cannot be sustained at any finite
temperature (i.e. $T_c=0$ for this model). It is reasonable to think that a
similar spin glass order occurs in our model for $J_0\neq 0$, provided
we are in the paramagnetic region $a_f<a<a_a$. Although the dynamical results
that will be presented below are clearly independent on any assumption regarding
the nature of the ground state, we will later conform to the idea
that a spin glass phase exists to
interpret the kinetic behaviors.

\subsubsection{Defective antiferromagnet ($a_a\le a\le1$)}\label{reg4}

In the region with $a>a_a$ there are very few and far apart ferromagnetic bonds.
Due to the condition (\ref{condition}), the couples of spins attached to these 
bonds will be aligned. This represents a defect in the otherwise perfectly
ordered antiferromagnetic state. 
Therefore in this region the system is an antiferromagnet with a fraction $1-a$ of isolated defects;
the name {\it defective antiferromagnet} is due to this.
One then has $m=0$ and 
$M\neq 0$ in this region, as it can be seen in Fig.~\ref{fig_magn}.
We see from Fig.~\ref{fig_gs_anti} that the local antiferromagnetic order parameter
$\sigma _i$ organises in large regions as the critical point $a_a$ is approached,
similarly to what $s_i$ does as the ferromagnetic transition at $a=a_f$ is narrowed.
This suggests that the size $\xi _a$ of such regions diverges at $a=a_a$, as it is 
sketched in Fig.~\ref{fig_substrate}, and that a continuous transition occurs.
 
\subsection{Equilibrium states at finite temperature} \label{eq_finiteT}

Since we will consider the evolution of a system quenched to $T_f>0$,
it is worth to discuss briefly the modifications to the above equilibrium 
picture at $T=0$ introduced by
a finite temperature. We must first keep in mind that the
{\it ferromagnetic-always-wins} condition (\ref{condition}) implies 
\be
J_+> |J_-|.
\label{newcond}
\ee 
For instance, with our choices of the parameters we have
$J_+=2.25$ and $|J_-|=0.25$.
In a standard system with
a definite value $J$ of the coupling constant it is $T_c\propto J$.
Although in the present model $J$ is not uniform,
upon raising the temperature one expects that
the antiferromagnetic ordering will be destroyed before the ferromagnetic one.  

Let us now consider the region with $a<a_f$, where ferromagnetic order
prevails.
In this region the critical temperature is expected to drop
from the Ising value $T_c(a=0)\simeq 2.269 J_+ \simeq 5.105$, to $T_c(a=a_f)=0$
(here and in the following temperature is measured in units of the Boltzmann constant).
Clearly, on the antiferromagnetic sector $a>a_a$ the corresponding 
critical temperature, that we will still denote with $T_c$, will 
drop from $T_c(a=1)\simeq 2.269 J_-\simeq 0.567$ to $T_c(a_a)=0$ upon decreasing $a$.

In the paramagnetic region, the spin glass phase is expected to be destroyed
by thermal fluctuations, no matter how small. Hence $T_c=0$ in this phase.
However, as we will discuss further below, the existence of spin glass order at
$T=0$ may strongly influence the dynamical properties at finite temperatures.

\section{Kinetics} \label{sec_kin}

The system is prepared in a fully disordered
initial state with uncorrelated spins pointing randomly up or down, 
corresponding to an equilibrium configuration at $T=\infty$,
and is then quenched at $t=0$ to a low final temperature $T_f$. 
We evolve the model by means of single spin flips governed by the Glauber
transition rates.

The main observable that we will consider is the inverse excess energy
\be
L(t)=[E(t)-E_\infty]^{-1},
\label{lt}
\ee
where $E_\infty$ is the energy of the equilibrium state at $T=T_f$. 
The latter has been obtained from the corresponding ground state by evolving it 
at $T=T_f$ until stationarity is achieved. For comparison, we have also found
the equilibrium state directly by means of parallel tempering techniques.
The values of $E_\infty$ found with the two methods are consistent.

When the system has a simple ferromagnetic or antiferromagnetic order, as in clean
or weakly disordered systems, the quantity
in Eq. (\ref{lt}) can be straightforwardly identified with the size of the growing ordered regions.
This is because in a coarsening process the interior of domains is in equilibrium and  
the excess energy is stored on the interface, so $E(t)- E(\infty)$ is proportional to their total length.
This in turn is given by the length of a single domain's boundary ($\propto L(t)^{d-1}$)
times the number of such domains ($\propto L^{-d}$), from which the
relation (\ref{lt}) between the size $L (t)$ of domains and the excess energy is obtained.

Notice that the above identification of $L(t)$ with an ordering length relies on a number
of assumptions: domains can be straightforwardly defined, they are equilibrated in their interior, and they
are compact objects (i.e. they have an euclidean dimension, not a fractal one).
While such assumptions are appropriate in a standard coarsening scenario,
they cannot apply to the present model for any choice of $a$. Notably, the identification
above almost surely fails in the paramagnetic region $a_f\le a \le a_a$.
In these cases $L(t)$ should prudentially be regarded simply as the inverse distance
from the equilibrium energy.

Of course, the typical size of the growing structures can be measured -- besides
as in Eq. (\ref{lt}), also in
many other ways. For example, in the ferromagnetic phase it can be easily extracted
from the equal time spin correlation function $\langle s_i(t)s_{i+r}(t)\rangle$,
where $i$ and $i+r$ are two sites at distance $r$.
Analogously, in a spin glass phase one might use the equal time overlap correlator
$\langle q_i(t) q_{i+r}(t) \rangle$, where 
\be
q_i=s_i(t)s_i^{GS}
\label{def_overlap}
\ee
is the overlap with the ground state.
However, already in the ferromagnetic region, the result need not to be
the same using different methods. Indeed, in a situation like the one of the ground state
at $a=0.4$ (upper right panel of Fig.~\ref{fig_gs}), for instance, there many
island with a finite extension. Then upon extracting a typical length from the correlation
function at late times one weights a lot the many small islands, which at
long times are already equilibrated (hence coarsening is interrupted in their interiors)
together with a comparatively smaller number of large islands. The latter are the only 
non-equilibrated regions within
which phase-ordering is still active. Instead, using Eq. (\ref{lt}) one only focuses on
those parts of the system where coarsening is still active, since inside the small
equilibrated islands $E(t)\equiv E_\infty$ by definition. Hence Eq. (\ref{lt}) is
more suited to qualify how phase-ordering proceeds in the regions where it is still at work,
while from the correlation function one obtains the average size of domains,
irrespective of their state, weather in equilibrium or not.
Since in this paper we are
more interested to address the dynamical mechanisms driving the kinetics,
we focus on the definition (\ref{lt}).

\begin{figure}[h]
\centering
\rotatebox{0}{\resizebox{.65\textwidth}{!}{\includegraphics{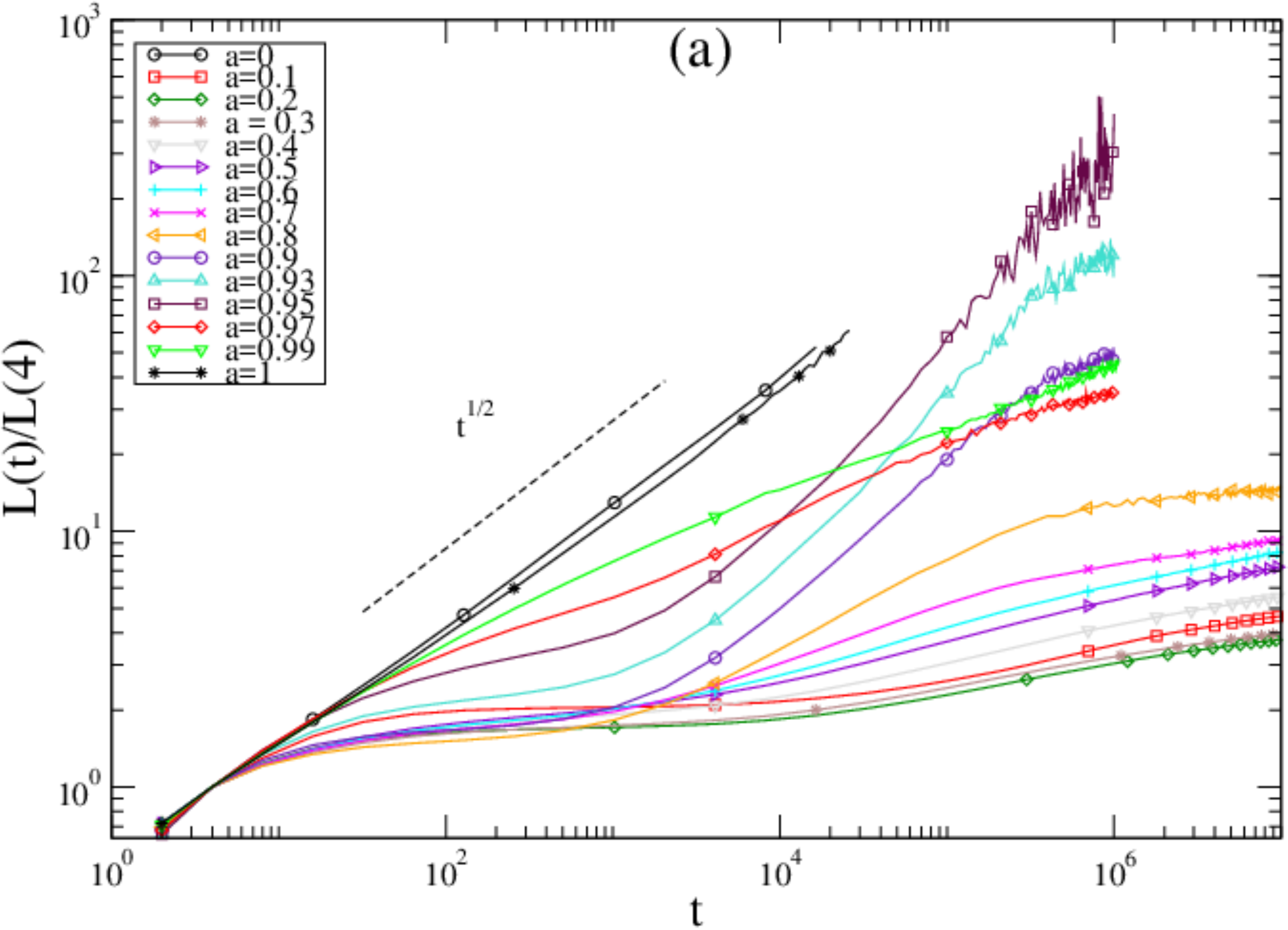}}}

\vspace{1cm}
\rotatebox{0}{\resizebox{.65\textwidth}{!}{\includegraphics{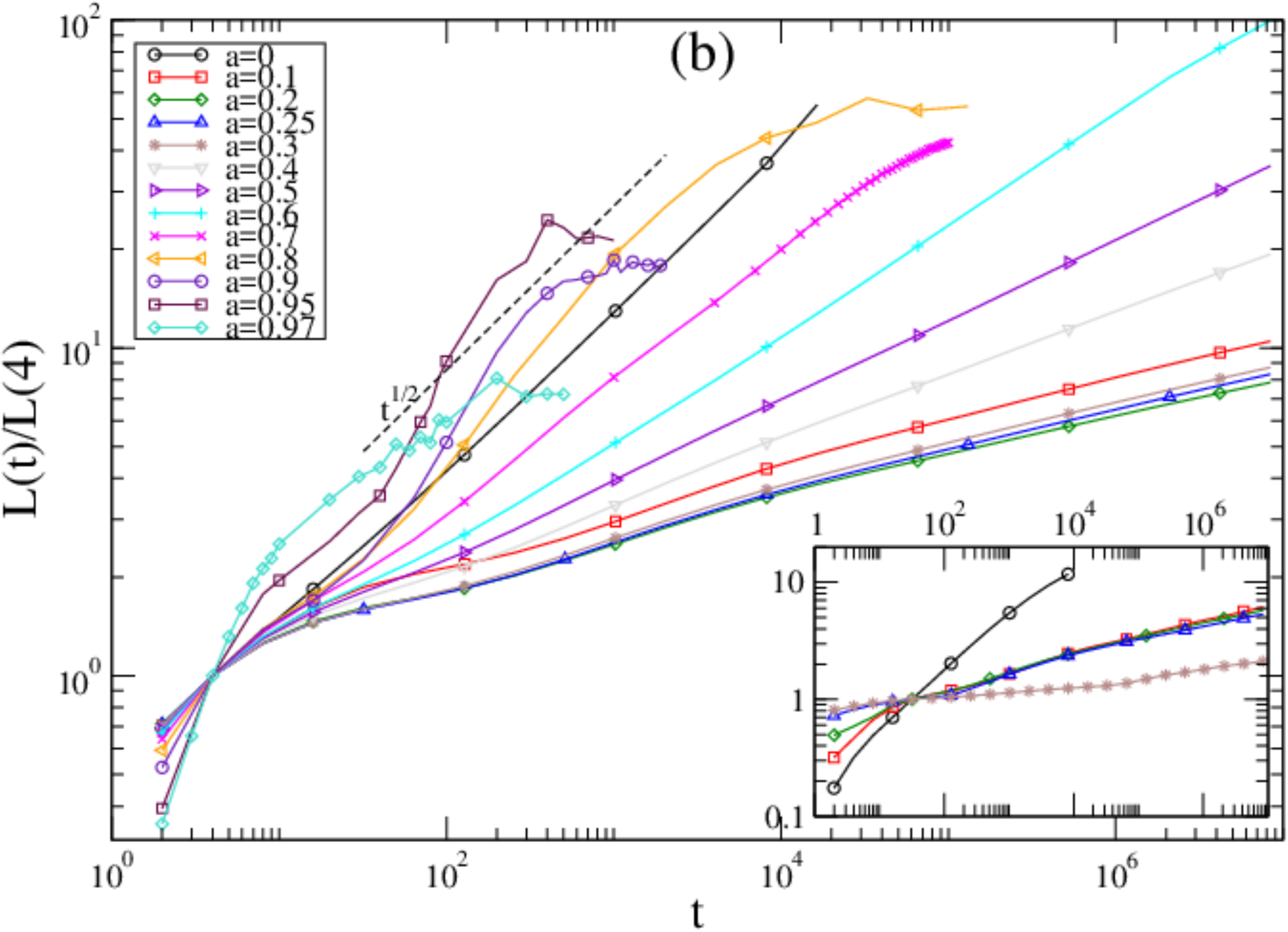}}}
\caption{$L(t)$ is plotted against time for a quench at $T_f=0.4$ ((a), upper part) and
at $T=0.75$ ((b), lower part),
for different values of $a$, in a double logarithmic plot. 
The black dashed line is the power-law $t^{1/2}$. The inset of the lower panel
shows, for a quench to $T_f$ and values of $a$ restricted to the ferromagnetic
region, the behavior of the characteristic length extracted from the spin correlation
function (see text).}.
\label{fig_quench}
\end{figure}

The difference between the value of $L$ obtained from the definition (\ref{lt})
or from the spin-spin correlation function
can be appreciated by looking to the inset of the lower
panel of Fig.~\ref{fig_quench}. Here, for a quench to $T_f=0.75$,
the typical length computed as in Eq. (\ref{lt}) is plotted for several values
of $a$ in the main picture, while the one obtained from the spin-spin correlation
is reported in the inset (only for values of $a$ in the ferromagnetic region).
The latter determination grows much slowly than the former at $a=0.4$, precisely 
because the ground states contains many small islands.

\begin{figure}[h]
\centering
\rotatebox{0}{\resizebox{.65\textwidth}{!}{\includegraphics{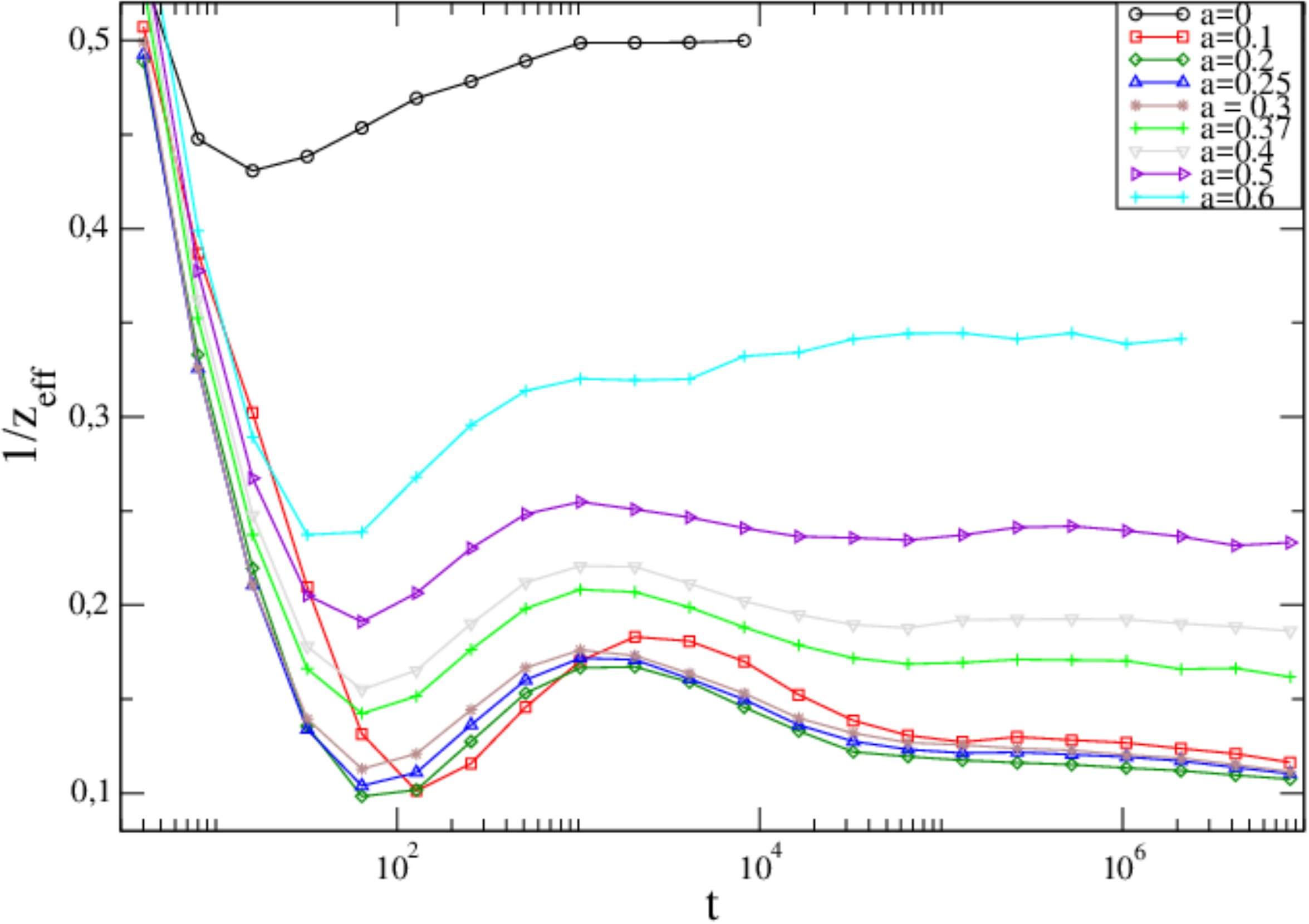}}}
\caption{The effective exponent $1/z_{eff}(t)$ is plotted against time for a
quench at $T=0.75$,
for different values of $a$, in a double logarithmic plot.} 
\label{fig_eff_exp}
\end{figure}

Let us now say a few words on the role of the final temperature $T_f$. 
Here we are interested in a situation where $T_f$ is very small. This is because usually
the kinetics of magnetic systems is more easily interpreted in this limit, and also because
low temperatures guarantees $T_f<T_c(a)$ -- the situation we are interested in -- in a wider
range of $a$ (see previous discussion about $T_c(a)$ in Sec. \ref{eq_finiteT}).
Setting $T_f$ to very small values, however, has the undesirable consequence that 
the kinetics becomes so sluggish that no appreciable
growth of $L(t)$ can be detected in the range of simulated times. 
In the following we will consider, out of many values $T_f$ used in the simulations,  
the two choices $T_f=0.4$ and $T_f=0.75$, that were found
to represent a good compromise between the two contrasting issues discussed above.
Notice that both these temperatures are much below the critical temperature $T_c(a=0)$ of the 
clean ferromagnet. On the other hand, while the former is smaller that of the clean antiferromagnet $T_c(a=1)$,
the latter is above. Let us stress that, in any case, since $T_c(a_f\le a\le a_a)=0$, for some values of $a$
the quench is necessarily made above the critical temperature. We will comment further below on 
the implications of this. 

\begin{figure}[h]
\vspace{2cm}
\centering
\rotatebox{0}{\resizebox{.95\textwidth}{!}{\includegraphics{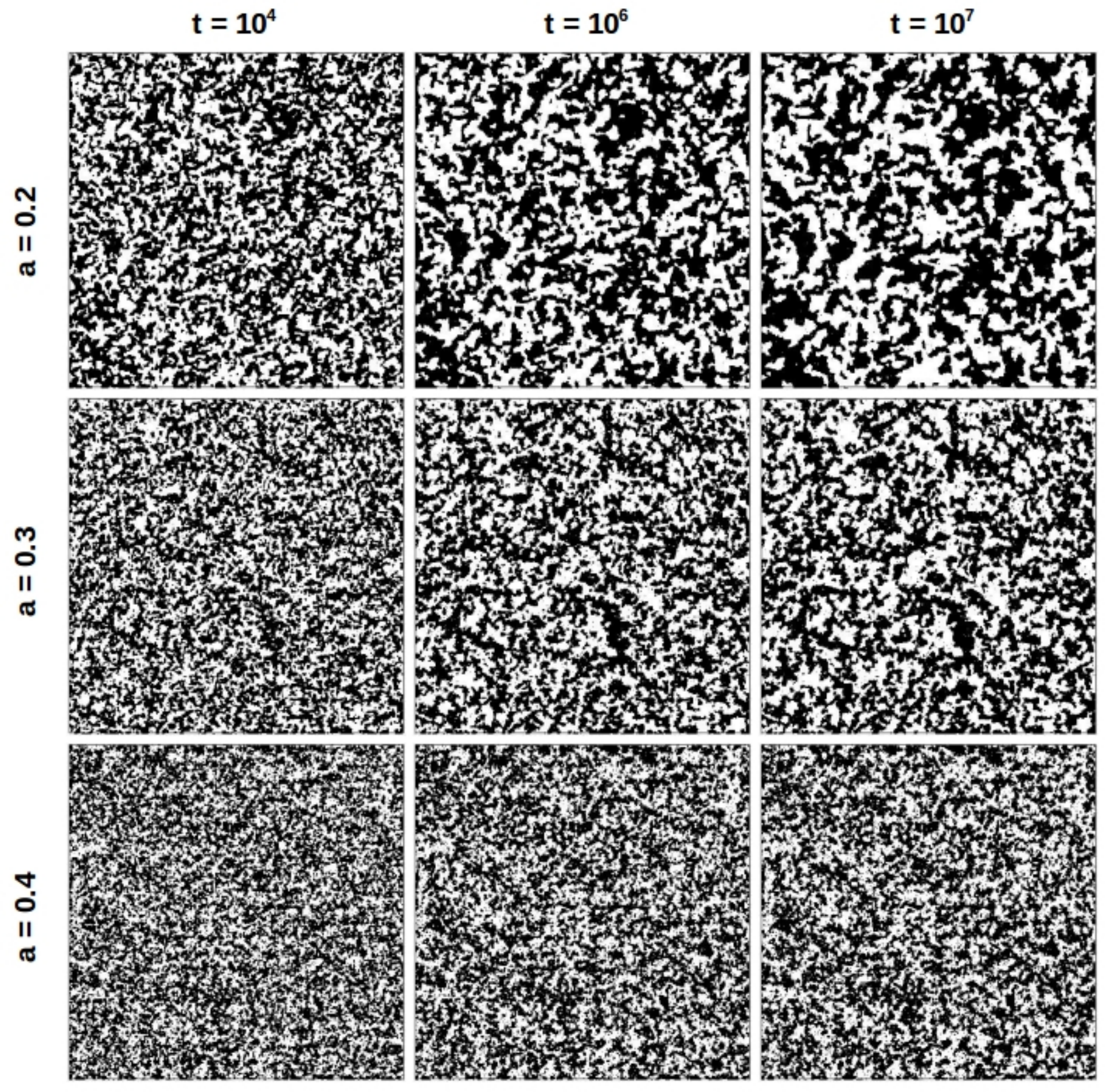}}}
\caption{Configurations of spins for a system of size ${\cal L}=512$ quenched
to $T_f=0.4$ for $a=0.2$ (upper row), $a=0.3$ (central row) and $a=0.4$ (lower raw). 
First columns report configurations at $t=10^4$, the central one at $t=10^6$, and the
left one at $t=10^7$.
Spins up are plotted in black, spins down in white.}
\label{fig_quench_ferro}
\end{figure}

\subsection{Simulations}

We now discuss the outcomes of our numerical simulations, splitting the
presentation in the following three section \ref{ferrodyn}, \ref{paradyn}
and \ref{antidyn} for quenches in the ferromagnetic, paramagnetic, and antiferromagnetic
region, respectively. The general behavior of $L(t)$, in the whole range
of values of $a$, is shown in
Fig.~\ref{fig_quench}, for $T_f=0.4$ and
$T_f=0.75$ (upper and lower panel, respectively).
Notice that we plot $L(t)/L(t=4)$ to better compare different curves.

As a general remark, we notice that at low temperature, such as for $T_f=0.4$,
$L(t)$ exhibit an oscillating behavior on top of the {\it neat} growth
(namely the kind of growth that one would have if such oscillations were smoothed out in some way).
This is quite commonly observed in
disordered or inhomogeneous systems at low $T_f$ \cite{noi_rb, noi_3d,parma13}
and is usually interpreted as due to the stop and go mechanism due to the pinning of interfaces.
Indeed, an interface gets trapped in the configurations where it passes through
weak bonds $J=J_-$.
For instance, the smallest energetic barrier $\Delta E$ encountered by a piece of interface 
occurs when it has to move from a position where it intersects a single antiferromagnetic bond
to one where it crosses only ferromagnetic ones. This situation is likely to be observed
when there are few antiferromagnetic bonds around, namely for small $a$.
In this case $\Delta E=J_+-J_-=2K$.
The associated Arrhenius time to escape the pinned state is $\tau \simeq \exp[-\Delta E/(K_BT)$.
With the parameters of our simulations one has $\tau \simeq  518$ for $T_f=0.4$ and 
$\tau \simeq 28$ for $T_f=0.75$. One sees in Fig.~\ref{fig_quench} that,
for $T_f=0.4$, this value is very 
well compatible with the time where $L(t)$, after becoming very slow, starts growing 
faster again (a rough agreement is found also for $T_f=0.75$, although in this case the oscillatory
phenomenon is only hinted).

Oscillations shadow the genuine
growth law, and it is therefore almost impossible to come up with any
quantitative statement about the neat growth, e.g. if it consistent with a power-law
or a logarithm or else.
As the final temperature is raised, the stop and go mechanism, although still present,
is less coherent, and the oscillations are smeared out. This is observed at $T_f=0.7$.
Moreover, the speed of ordering increases upon raising $T_f$, as it is expected
in the presence of activated dynamics. 

\subsubsection{Ferromagnetic region $0\le a\le a_f$.} \label{ferrodyn}

Snapshots of the system's configuration at different times after a quench to $T_f=0.4$,
and for three
different choices of $a$, are shown in Fig.~\ref{fig_quench_ferro}.
For any value of $a$ one clearly observes a coarsening phenomenon with
domains of the two phases growing in a self-similar way in time.
Upon increasing $a$, domains look more jagged and indented, presumably due
to the nearing of the critical point at $a=a_f$ where, as usual in second order
phase-transitions, a fractal structure is expected to appear.

Concerning the size $L(t)$ of such growing structures (Fig. \ref{fig_quench}),
starting from the pure case with $a=0$ (black curve with circles), where the expected
behavior $L(t)\propto t^{1/2}$ is well represented, the growth law slows 
down upon rising $a$, but this occurs only up to a certain value $a$, which 
we interpret as $a=a^*$ and is located around $a=0.2$. 
Upon increasing $a$ beyond $a^*$, the phase-ordering
process speeds up again, up to $a_f$. This is very well observed both for $T_f=0.4$
and $T_f=0.75$. The value of $a^*$, at these two temperatures, is comparable, as it 
is expected since this quantity was previously defined in a purely geometrical way.

In previous papers~\cite{noi_diluted,noi_rb,noi_3d} we have shown that the growth rate $r=dL/dt$
of $L(t)$, namely the speed of growth, varies in a non-trivial way as the amount of 
disorder in a ferromagnetic model is changed.  Specifically, in a system where a fraction
$d$ of lattice sites or bonds is randomly removed,
the speed of growth is at its maximum for the pure system at $d=0$. As $d$ is
increased, disorder pins the interfaces and $r$ decreases. The larger is $d$, the larger
is the pinning effect, and the smaller is $r$. However this is only true
for moderate values of $d$, namely for $d\le d^*$. Raising $d$ above $d^*$ one observes
that $r$ starts to increase again up to the largest possible value of $d$, namely 
$d=d_p$, where $d_p=1-p_c$,  $p_c$ being the critical threshold of random percolation.
For $d>d_p$ the network where the spins live becomes disconnected and the ferromagnetic
properties are lost. It turns out that $d^*$ is roughly in the middle between $d=0$ and $d=d_p$.
The interpretation of this non monotonic behavior of $r$, given
in~\cite{noi_diluted,noi_rb,noi_3d},
is the following: besides the pure case, where the speed of growth is at its maximum,
the other network where $L(t)$ grows relatively fast (but slower than in the pure case)
is the percolation fractal at $d=d_p$. This is argued 
to be due to the critical properties of such network, which in turn are responsible of the
fact that, right at $d=d_p$, the transition temperature of the model
$T_c(d=d_p)$ vanishes. Indeed the fractal properties of the structure at $d=d_p$ soften
the energetic barriers that pin the interfaces, making in this way the evolution faster.
Since there are
a global and a relative maximum of $r$ at $d=0$ and $d=d_p$, respectively, 
there must be a minimum somewhere in between. This value is $d^*$. This explains why
$r$ in a non monotonic function of $r$. 

Also in the present model the geometric properties of the system are such that 
there is a point, namely $a=a_f$, where $T_c(a=a_f)$ vanishes
(actually the same occurs at $a=a_a$, where an analogous
discussion is expected to hold). Therefore, upon repeating the argument above
(with the obvious replacements $d\to a$ and $d_p\to a_f$),
one could expect $r$ to be at its global maximum at $a=0$,
to decreases as $a$ increases up to a value 
$a^*$ located in between $a=0$ and $a=a_f$, and then
to rise again up to $a=a_f$. This is precisely what we see in
Fig.~\ref{fig_quench}.
The non-monotonic dependence of the speed of growth
on disorder, therefore, qualifies as a
rather general property of ferromagnetic systems, and
a common interpretation for different models can be provided.

Let us also mention that in~\cite{noi_diluted,noi_rb,noi_3d} not only
the non monotonic behavior discussed insofar was shown, but a quantitative
conjecture was put forth: the asymptotic growth law should 
be of the logarithmic type in the whole disordered region $0<d<d_p$,
while it ought to be algebraic $L(t)\propto t^{1/z}$ both in the clean case $d=0$
(with $z=2$) and at $d=d_p$ (with $z>2$ and
$T_f$-dependent). For $d$ close to $d=0$ (or to $d_p$) the logarithmic
growth is shadowed preasymptotically by the algebraic behavior induced
by the proximity of the clean point $d=0$ (or the percolative one $d=d_p$).
Notice that, in some cases, algebraic preasymptotic behaviors
have been shown to leave room to a truly logarithmic growth only after
huge times, a notable example being the random bond
Ising model with a continuous distribution of ferromagnetic coupling constants
\cite{noi_rbpowerlog}. This originated a lot of contradicting conclusions
in the past.

If a mechanism akin to the one discussed above is at work also in the present model
one would expect
an asymptotic logarithmic behavior (after a -- possibly slow -- crossover) for any 
$0<a<a_f$, and a power-law behavior of $L(t)$ right at $a=a_f$.
While nothing can be said, as discussed above, for quenches to $T_f=0.4$ due to the
oscillating nature of the curves, this conjecture can be tested to some extent
in the data for $T_f=0.75$.
This can be done by computing the effective exponent $z_{eff}$ defined as
\be
\frac{1}{z_{eff}(t)}=\frac{d[\ln L(t)]}{d[\ln t]}.
\ee
This quantity is plotted in Fig.~\ref{fig_eff_exp}.
For $a=0$ it approaches the expected asymptotic value $1/z_{eff}=1/2$
starting from relatively early times $t\simeq 10^3$. As $a$ is progressively increased
from $a=0$ to $a=0.4\simeq a_f$, the effective exponent becomes, for a given
late time, initially smaller (in the range $0\le a\le a^*\simeq 0.2$) and then rises
again (moving $a$ in the range $a^*\le a\le a_f$).

Concerning the time evolution of $1/z_{eff}$ our data clearly show that it
keeps steadily decreasing in the late regime with $t\gtrsim 10^4$ for all the values
of $a$ in the range $0<a\le 0.3$. The decrease is rather slow but reliable.
This implies that the growth law of $L(t)$ is slower than algebraic. 
Although the curves for $L(t)$ span a vertical range which is too limited to
allow a precise determination of such law
(furthermore, a weak oscillation is present up to times of order $10^4$),
we can at least conclude that the decrease of $1/z_{eff}$
agrees with the expectation of a logarithmic behavior.
Data for $a=0.4\simeq a_f$, on the other hand, are quite well consistent with a
constant behavior of $1/z_{eff}\simeq 0.19$ at late times, and this also agrees
with the conjecture discussed above. Finally, the effective exponent
looks rather constant also for $a=0.37$. This can be ascribed to the preasymptotic
algebraic behavior induced by the proximity of the percolation point $a=a_f$.
We expect, therefore, that a decrease of $1/z_{eff}$ would be observed also for
$a=0.37$ if sufficiently long times could be accessed in the simulations.

Notice that an algebraic law is also observed in the whole paramagnetic region,
an interpretation of which will be provided in the next section.
Therefore, the different asymptotic behavior -- i.e. algebraic versus logarithm --
observed at $a=a_f$ with respect to the rest of the ferromagnetic region 
$0<a<a_f$ can also be interpreted upon thinking $a_f$ as the lower limit of the
paramagnetic region where algebraic behaviors are observed. We will comment further
on this point below.

\subsubsection{Paramagnetic region $a_f<a<a_a$} \label{paradyn}

\begin{figure}[h]
\vspace{2cm}
\centering
\rotatebox{0}{\resizebox{.95\textwidth}{!}{\includegraphics{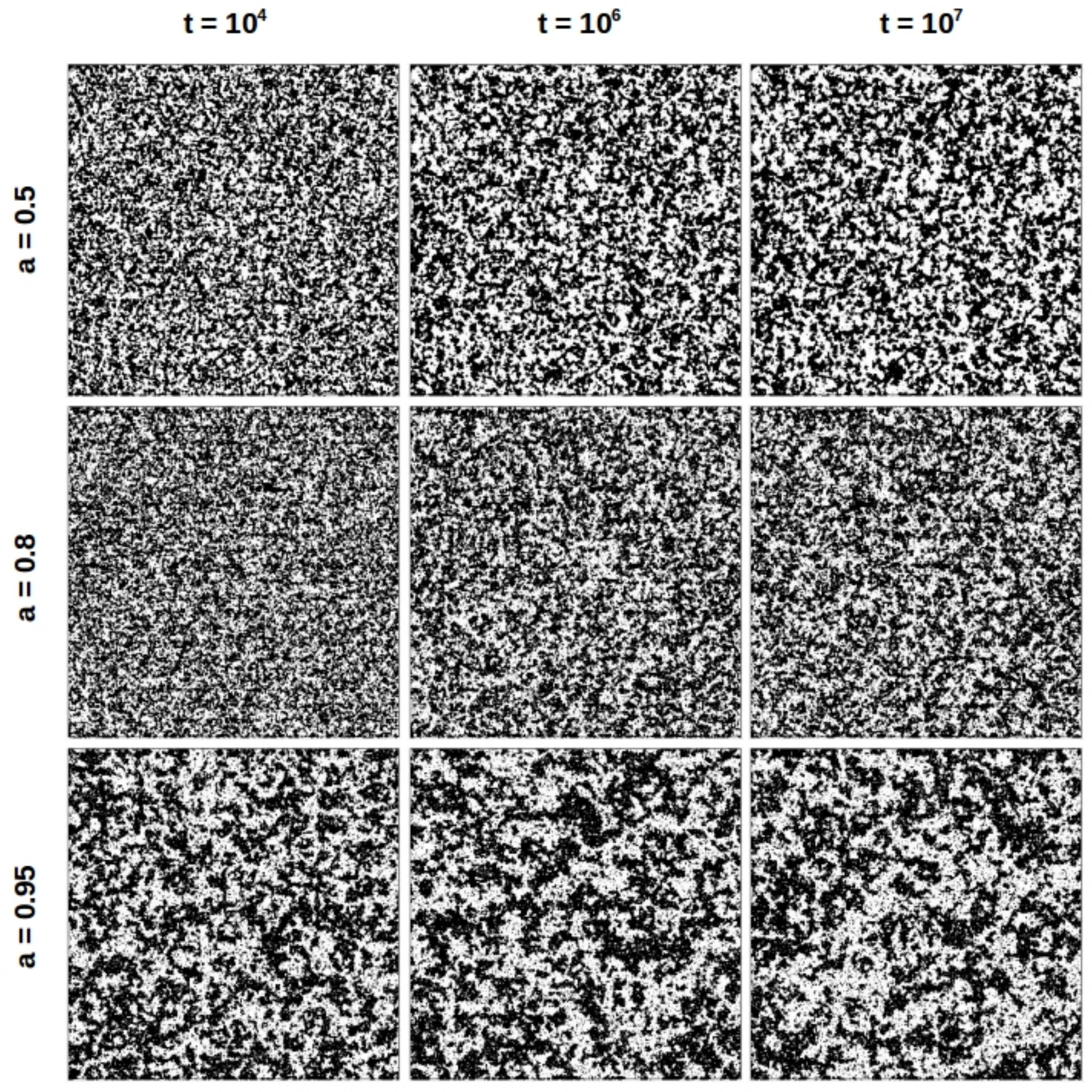}}}
\caption{Configurations of the overlap, as defined in Eq. (\ref{def_overlap}))
for a system of size ${\cal L}=512$ quenched
to $T_f=0.4$ for $a=0.5$ (upper row), $a=0.8$ (central row) and $a=0.95$ (lower raw). 
First columns report configurations at $t=10^4$, the central one at $t=10^6$, and the
left one at $t=10^7$.
Positive overlaps are plotted in black, negative ones in white.}
\label{fig_quench_para}
\end{figure}

In two-dimensional spin glasses (corresponding to $J_0=0$ in our model)
it was shown~\cite{agingint,sgpower1,sgpower2}, that
the existence of a spin glass phase at $T=0$ rules the
kinetic in a long lasting preasymptotic regime.
The situation is akin to what observed in the clean Ising model at the lower
critical dimension $d_L=1$ after a quench to a finite but low enough
temperature. In this case it can be shown~\cite{Glauber,noi_onthe} that the kinetics
proceeds as in a quench to $T_f=0$ until $L(t)\simeq \xi$, where
$\xi =e^{2J_0/T_f}$ is the equilibrium coherence length. Since this quantity is
huge, the preasymptotic regime where non equilibrium kinetics is observed extends
to very long times at low $T_f$. Afterwards, aging is interrupted and
the system equilibrates. In the $d=2$ spin glass it is found that,
as long as the preasymptotic stage is considered,
a growing length can be identified which exhibits an algebraic behavior~\cite{sgpower1}.

In the present model, assuming that a kind of spin glass order is developed
at $T_f=0$ in the paramagnetic region $a_f<a<a_a$, we expect to observe
a similar phenomenon. This can be already appreciated at a qualitative level in
Fig.~\ref{fig_quench_para}. In this set of figures we plot the local overlap
(Eq. (\ref{def_overlap}))
between the actual dynamical spin configurations and the ground state,
for different times after a quench to $T_f=0.4$, and for three
different choices of $a$. 
Interestingly, also in this paramagnetic phase, for any value of $a$,
one clearly observes a coarsening phenomenon with
domains of the two phases growing in a self-similar way in time, although
quite slowly. This agrees with what observed in~\cite{sgpower1}.
Moreover, there is no signal of equilibration at any time,
nor the growth seems to be interrupted. As a further comment, we notice that
configurations appear much more rugged for larger values of $a$.

The study of the approach to equilibrium can be made more quantitative by inspection of
Fig.~\ref{fig_quench}, where $L(t)$ for the various cases is plotted.
Here one sees that, as already anticipated, data for $T_f=0.75$ are consistent
with an algebraic growth $L(t)\propto t^{1/z}$, with an $a$-dependent exponent,
in agreement with~\cite{sgpower1}.
Data for $a\ge 0.7$ clearly bend downwards at late times,
indicating that equilibration is starting to be achieved.
The algebraic increase of $L(t)$  is further confirmed by inspection of
the effective exponent $z_{eff}$ in Fig.~\ref{fig_eff_exp}.
This quantity stays basically constant, besides some noisy behavior, in the
late time regime $t\gtrsim 10^3-10^4$.
Notice also that $1/z_{eff}$ raises as $a$ is increased.
This can be ascribed, at least partly, 
to the fact that the largest barriers encountered are associated to the positive
couplings (since condition (\ref{newcond}) holds), and the number of the latter
is reduced upon increasing $a$. 

A power-law for $L(t)$ in this paramagnetic region,
as opposed to the logarithmic one in most
disordered ferromagnetic models, including the one at hand for $0<a<a_f$,
can perhaps be read
into the different structure of the low energy states.
For a ferromagnet there are two degenerate ground states
separated by an energetic barrier. A common picture of a frustrated system is,
instead, one with many quasi-equivalent low-energy states.
Taking advantage of entropic effects, the
system can move among these states lowering in this way
the free energy barriers. This could speed up the evolution from
logarithmic to algebraic.

\subsubsection{Region with antiferromagnetic order $a\ge a_a$.} \label{antidyn}

\begin{figure}[h]
\vspace{2cm}
\centering
\rotatebox{0}{\resizebox{.95\textwidth}{!}{\includegraphics{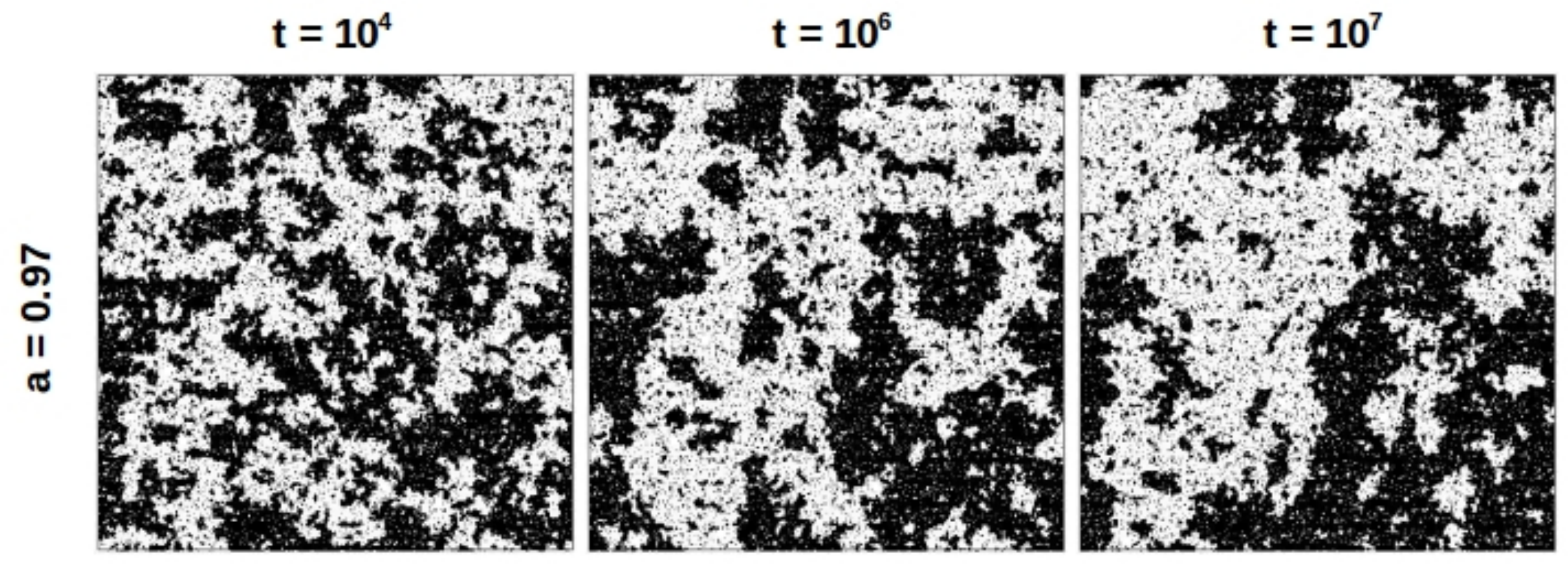}}}
\caption{Configurations of the staggered spin, as defined in Eq. (\ref{stagspin})
for a system of size ${\cal L}=512$ quenched
to $T_f=0.4$ for $a=0.97$ for times $t=10^4$, $t=10^6$, $t=10^7$.
Positive $\sigma _i$'s are plotted in black, negative ones in white.}
\label{fig_quench_anti}
\end{figure}

In Fig.~\ref{fig_quench_anti} we plot the configurations of the staggered spin
$\sigma _i$ -- see Eq. (\ref{stagspin}) -- at different times, which clearly show
the coarsening phenomenon. In this region with antiferromagnetic order we expect a
situation mirroring the one discussed in the ferromagnetic region, with the obvious
correspondences $a=0 \leftrightarrow a=1$, and $a=a_f \leftrightarrow a=a_a$.
Since in this case $T_c(a=1)\simeq 0.567$ as already discussed, there is no room
left to observe coarsening for $T_f=0.75$ for any value of $a<1$ (indeed for $T_f=0.75$
-- in the lower panel of Fig.~\ref{fig_quench} -- $L(t)$ flattens very soon,
saturating to the equilibrium value).
Let us then focus only on the upper panel of Fig.~\ref{fig_quench}, where
data for $T_f=0.4$ are presented. Here one observes the non-monotonic behavior of the
speed of ordering $r$ already discussed when considering the ferromagnetic phase.
Upon decreasing $a$ from the clean value $a=1$ the kinetics becomes very soon much slower
in going to $a=0.97$, and then $r$ increases again until the border of 
the paramagnetic phase is touched at $a=a_a\gtrsim 0.95$.

\section{Conclusions}

In this paper we have studied numerically a two-dimensional random bond Ising model
where a fraction $1-a$ of coupling constants is positive $J=J_+>0$ and the remaining ones
are negative, $J=J_-<0$. The choice of $J_+$ and $J_-$ has been made according to a
{\it ferromagnetic-always-wins} condition which strongly favors ferromagnetic ordering.
We have classified the low temperature equilibrium states
according to the values taken by the magnetisation $m$ and the staggered magnetisation
$M$. We have shown the existence of a ferromagnetic and an antiferromagnetic
phase for $a<a_f$ and $a>a_a$. In between, for $a_f<a<a_a$, there is
a paramagnetic phase at any temperature $T$, presumably with spin glass order at $T=0$.

The main focus of the paper has been on the off-equilibrium evolution of the model after
a quench from an infinite temperature disordered state to low temperatures.
In the ferromagnetic and antiferromagnetic phases the process amounts to
the much studied coarsening phenomenon in the presence of quenched disorder,
which is characterised by the ever-lasting increase of the typical domains' size $L(t)$.

Also in the present model we find that the speed of the non equilibrium evolution
varies in a non monotonic way as the amount of disorder $a$ is increased,
similarly to what observed in other disordered ferromagnets~\cite{noi_diluted,noi_rb,noi_3d}.
Specifically there exists a value $a^*\sim 0.2$ where the kinetics is slower
than for any other value of $a$. We have also been able to show that the
growth law of $L(t)$ is slower than algebraic, i.e. of a logarithmic type, in the whole
ferromagnetic (and antiferromagnetic) region $0<a<a_f$ (or $a_a<a<1$).
Interestingly enough, this is true both for $0\le a^*$ and for
$a^*\le a<a_f$, irrespectively of the fact that 
the structure of the ferromagnetic state is different in these two sectors, because
the system is
a perfect ferromagnet (i.e. $m^2=1$ at $T=0$) in the former range while
it contains a number of defects (antiferromagnetic inclusions) in the latter
(so that $m^2<1$ at $T=0$).

Similar considerations can be made in the paramagnetic region. Here we find that
$L(t)$, the inverse excess energy, grows algebraically in the whole phase, irrespectively
of the fact that the geometrical properties of the bond network greatly change
as $a$ is varied in $[a_f,a_a]$. 

These results seem to indicate that $L(t)$ is able to discern between
systems with a ferromagnetic phase extending below a finite critical temperature
$T_c>0$ from the others. In the former $L(t)$ grows in a logarithmic way, whereas
a power-law is observed otherwise. The results of this paper show
that this property, which was already found
in models of disordered magnets without frustration~\cite{noi_diluted,noi_rb,noi_3d},
extends its validity to the realm of frustrated systems, pointing towards a
general robustness.

\acknowledgments

We thank Hamid Khoshbakht for discussions and significant help in the determination of
the ground states.
We thank Federico Ricci-Tersenghi for useful discussions.
F.C. acknowledges financial support by MURST
PRIN 2015K7KK8L.

\appendix

\section{Suppression of ferromagnetic order for $a\gtrsim a_f$} \label{app1}

\begin{figure}[h]
\vspace{2cm}
\centering
\rotatebox{0}{\resizebox{.44\textwidth}{!}{\includegraphics{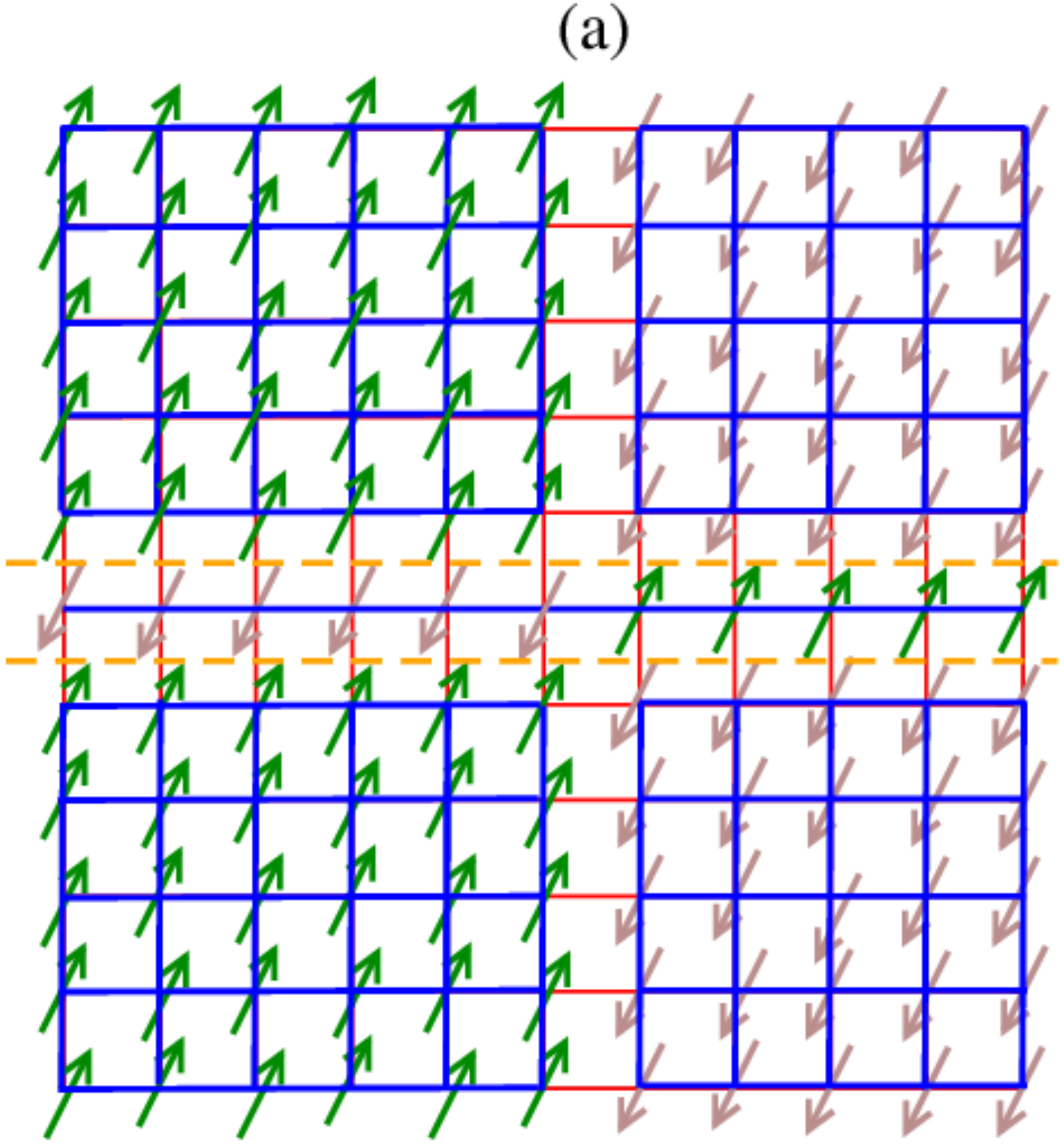}}}
\hspace{2cm}
\rotatebox{0}{\resizebox{.42\textwidth}{!}{\includegraphics{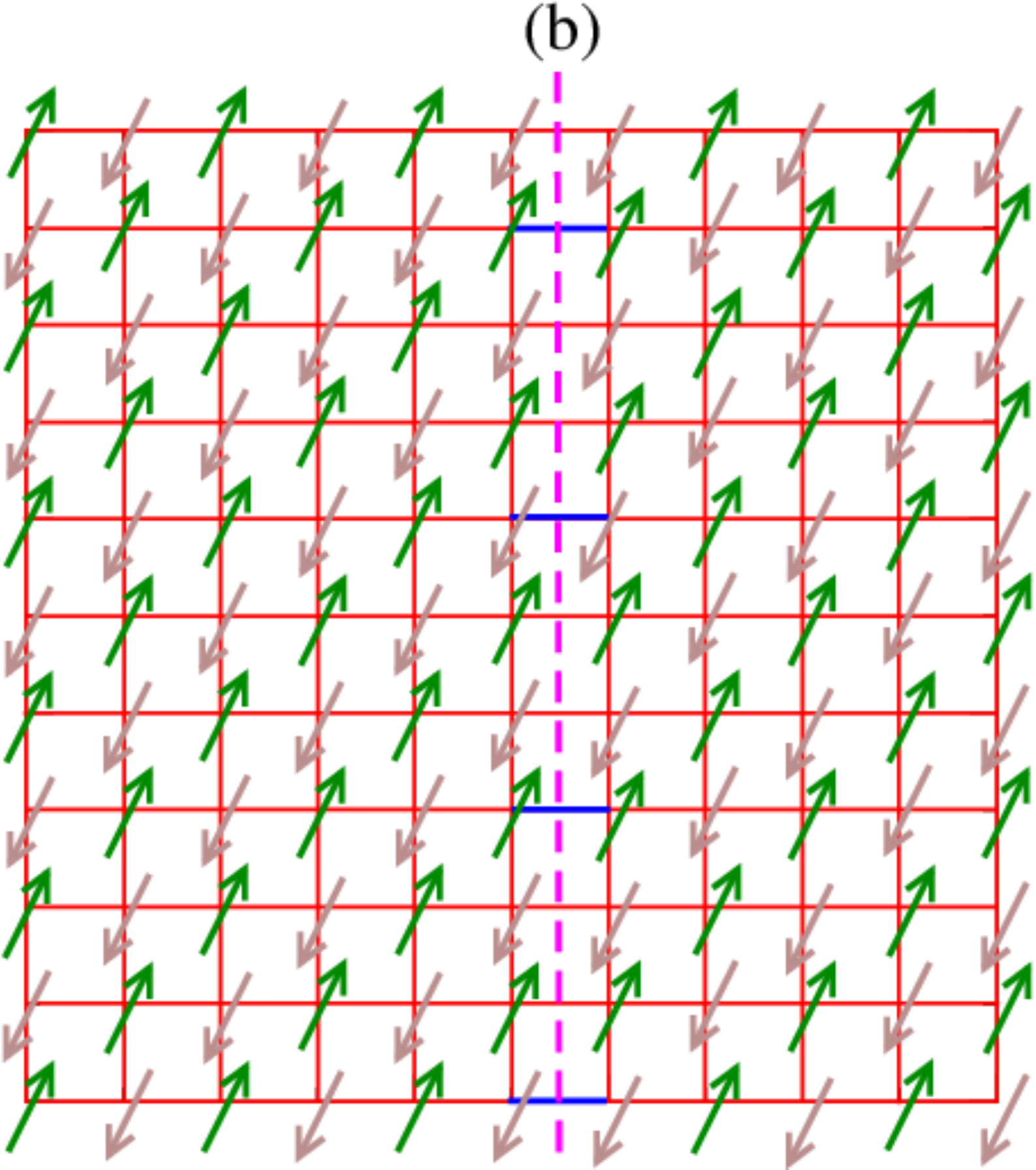}}}
\caption{Schematic representations of the ground states of the system in the regions
$a_f<a<a_p$ ((a), left part) and $a_p<a<a_a$ ((b), right partl). 
Antiferromagnetic and ferromagnetic bonds are drawn in red and
blue, respectively. Spins up and down are colored in green and brown.
In the left panel, the two dashed orange lines indicate paths of antiferromagnetic bonds.
In the right panel, the dashed magenta line is an interface in the antiferromagnetic
order.}
\label{fig_para}
\end{figure}

For $a\gtrsim a_f$ the amount of negative bonds is sufficient to
spoil the ferromagnetic order. This may happen, for instance,
when the spanning sea of ferromagnetic bonds have
a thin part,
like the horizontal path within the two dashed orange lines 
in the left panel of the schematic Fig.~\ref{fig_para}, 
along which spins cannot keep the same orientation without increasing
the total energy.
In the situation sketched it is
easy to check that the represented configuration,
minimizes the energy.

This picture has been presented to easily grasp the
properties of the ground state, but it is not appropriate to describe
the situation with $a<a_p$. Indeed for such values of $a$ there cannot be a spanning
path of antiferromagnetic bonds, while it is present in Fig.~\ref{fig_para} (along the dashed orange lines).
However one can easily check that the ground state does not change if a certain fraction
$f\le 1/z$ (in this case $f\le 1/4$)
of the antiferromagnetic bonds crossing the dashed orange lines are turned into ferromagnetic 
ones. In this new situation there are no spanning clusters of antiferromagnetic bonds, but
the ground state is still split into four pieces with discording magnetization. 

A computation of $a_f$ looks very difficult, since this amount to
evaluate the smallest probability $a$ such that a spanning object formed by
a fraction $1-f$ of antiferromagnetic bonds exists.
An (admittedly very rough) estimation is the following:
We know that at $a=a_p$ a spanning path of antiferromagnetic bonds exists.
If the probability is decreased to $(1-f) a_p$, a fraction $f$ of such
antiferromagnetic bonds will be converted to ferromagnetic ones.
This provides $a_f\simeq (1-f)a_p$. In our two-dimensional case this yields
$a_f\simeq 3/8=0.375$, to be compared with the observed value
$a_f\gtrsim 0.4$.

\section{Suppression of antiferromagnetic order for $a\lesssim a_a$} \label{app2}

The reason why antiferromagnetic order cannot establish up to fractions of
negative bonds as large as $a_a\simeq 0.95$
can be easily understood by looking at the schematic ground state situation 
of the right panel of Fig.~\ref{fig_para}. In this picture it is seen that also a small amount 
of ferromagnetic bonds can induce an interface (of the antiferromagnetic
type -- marked by a dashed magenta line) in the system. Indeed it is easy to check that, due to the condition (\ref{condition}),
the situation where such an interface is removed corresponds to an higher energy.
Because of the interface the system is split in regions with mismatched staggered magnetization and hence $M=0$. 
Clearly, the formation of the interface in the right panel of Fig.~\ref{fig_para} becomes
energetically unfavorable if the density of ferromagnetic bonds 
(basically the distance between blue bonds in Fig.~\ref{fig_para})
becomes too small. A rough estimation of the value of $a$ where this occurs
is the following. The average distance 
between ferromagnetic bonds is $\lambda _f$. Along an interface, such as the one
plotted in Fig.~\ref{fig_para}, there is a ferromagnetic bond of strength $J_+$
every $\lambda _f$ antiferromagnetic ones (of strength $J_-$).
Hence the interface cannot be sustained if $\lambda _f |J_-|>J_+$, namely for
\be
(1-a)^{-d}>\frac{J_+}{|J_-|},
\label{condanti2}
\ee
where we have used Eq. (\ref{lambdaf}).
With the choice $J_0=1$, $K =1.25$ adopted in our simulations this
would predict the interface instability at $a=a_a=0.7$.
This is only a lower bound to the value of $a_a$, since it is clear that
besides having condition (\ref{condanti2}) obeyed, other conditions must apply.
For instance an antiferromagnetic bond must be guaranteed next to the ferromagnetic
ones. Indeed we see in Fig.~\ref{fig_magn} 
that the paramagnetic phase extends much beyond $a=0.7$,
at least up to $a_a\sim 0.95$.

\end{document}